\documentstyle[epsf,12pt]{article}
\newlength{\dinwidth}
\newlength{\dinmargin}
\begin{document}

\begin{center}{\Large{\bf

Remarks Concerning the Study of Four-Jet Events from Hadronic Decays
of the~Z$^0$. 
}}
\end{center}

H.Jeremie\footnote{e-mail: jeremie@lps.umontreal.ca}, P.Leblanc, E.Lefebvre, 
D.Davignon, G.Karapetian;~~~Universit\'e 
de Montr\'eal\footnote{ this is the English version of
an article which appeared in French in ref.~\cite{CANJOUR}}.

{\large \underline{Abstract}}: The angular correlations of four-jet 
events from hadronic decays of the
Z$^0$ have been studied in the past mainly to extract from them the
fundamental constants of quantum chromodynamics called colour factors.
Previous studies have used all the available phase-space in
order to maximize statistics. In this note we want
to point out the possibility that significant differences between
experiment and theory in restricted regions of phase-space 
might have escaped detection. Such differences could be a
harbinger of the existence of new particles. Some preliminary results are 
presented.


\section{Introduction }

In second-order QCD perturbation theory four-jet events arise 
from the production of four
hard partons and their subsequent fragmentation. The cross-section
for four-parton production by electron-positron annihilation can then be 
written as:

\begin{equation}
\frac{{\mathrm{d}}\sigma(x)_{\mathrm{quatre-partons}}}{{\mathrm{d}}x}\;=f_{\mathrm{q}} D_{\mathrm{q\overline{q}q\overline{q}}}(x) + (1-f_{\mathrm{q}}) D_{\mathrm{q\overline{q}gg}}(x)
\end{equation}

where $D_{\mathrm{q\overline{q}q\overline{q}}}(x)$ and
 $D_{\mathrm{q\overline{q}gg}}(x)$ are differential, normalized, theoretical distributions of variables
$x$ for events whose final state consists of four quarks, and events consisting of two quarks and two gluons respectively
\footnote{ $D_{\mathrm{q\overline{q}gg}}(x)$ contains events where both quarks radiate a gluon as well as events where
one radiated gluon splits into two gluons.},
while ${\mathrm{d}}\sigma(x)_{four-parton}$/${\mathrm{d}}x$ is the normalized distribution of the data.
The value for the fraction $f_{\mathrm{q}}$ is a fundamental prediction of QCD,
it is approximately proportional to $T_R$, one of the QCD colour factors.
It can be extracted from a sample of four-jet events by varying
$f_{\mathrm{q}}$ to obtain a best fit to a measured distribution, 
if $D_{\mathrm{q\overline{q}q\overline{q}}}(x)$ and 
$D_{\mathrm{q\overline{q}gg}}(x)$ are sufficiently different from each other
(see f.ex. figure~\ref{hep_chibz}d) 
Such differences arise because in the case of ${\mathrm{q\overline{q}q\overline{q}}}$ events the primary
quark-antiquark pair radiates a polarized gluon which splits into two quarks of spin 1/2, while 
in the ${\mathrm{q\overline{q}gg}}$
case this intermediate gluon splits into two gluons of spin one. Angular momentum conservation
then requires the distribution of the final quark-antiquark pair with respect to the initial pair
to be different from that where the final partons are gluons.  
For practical reasons we preferred to work directly with $f_{\mathrm{q}}$ 
than with
$T_R$, the two approaches being equivalent.

If the intermediary gluon would have the possibility to split into fermions
other than those with the canonical QCD flavours, the value of 
  $f_{\mathrm{q}}$ (or  $T_R$) would show an increase, constituting thus
a possible signal of some new physics  \cite{GLUIN}.

The main theoretical distributions we used here  were obtained
from the  second-order (O($\alpha_s^2$)) QCD matrix element calculation by
Ellis, Ross and Terrano (called ERT model hereafter) \cite{ERT}, as implemented
in the JETSET simulation package \cite{JETST}, which includes
hadronization of the four partons. 
The parameters of the model were adjusted so as to reproduce energy-energy 
correlations and event-shape distributions
measured by OPAL~\cite{EECORR}~\footnote{
The values of the fragmentation  parameters we used were: 
PARJ(21)=0.49, PARJ(41)=1.8, PARJ(42)=0.6. }, and the scaled invariant mass
cut-off of the ERT four-parton generator was $y_{\mathrm{min}}$=0.01. 

Higher order QCD processes will also contribute to the production of four-jet events, 
in which case the value of $f_{\mathrm{q}}$ derived  from eq.(1) will no longer yield a correct measurement of 
this quantity.  However, $f_{\mathrm{q}}$ still provides a useful method of comparison between data and models. 
Since this method relies only on comparisons of shapes of angular
correlations and no actual four-quark events are ever identified,
we call the measured quantity the ``apparent'' fraction of four quark
events.

To gauge the influence of higher orders, 
a hybrid model combining matrix elements and parton 
showers, and a matrix element model generating pure five-parton events 
were also investigated 
(see refs.\cite{ANDRE} and \cite{ZEPP}).
Such a hybrid model is expected to simulate a number of higher order
effects, while the 5-parton events are one of the important contributions
of next-to-leading order(NLO) calculations.

Using all events produced in the entire region of phace-space (except for
a threshold), previous
investigations  (refs. \cite{ALEPH} to \cite{ALEPH3}) found good agreement
with QCD predictions for the overall global value of $T_R$, equivalent 
to  $f_{\mathrm{q}}$.  

As previously mentioned, in this work we will try to obtain more information
by examining the evolution of  $f_{\mathrm{q}}$ 
as a function of variables which permit a subdivision of the phase-space.
We chose the variable $m_3+m_4$, the sum of the masses of the two least 
energetic jets.

\section{Experimental Method}
\subsection{Event Selection}

The events were recorded with the OPAL detector \cite{OPALDET} at 
LEP{\footnote{Although data taken with the OPAL detector were used,
the analysis itself was done independently from the OPAL collaboration. 
Only the authors cited above assume responsibility for this analysis.},
details of  the trigger selection and 
on-line 
filtering system can be found in refs. \cite{XX} and \cite{YY}.
A standard selection of multihadronic events was applied \cite{ZZ}.
For the present analysis charged particles, measured in the central tracking system of the
OPAL detector, and
clusters formed by  showers in the electromagnetic and hadronic
calorimeter were used. Additional corrections were applied to minimize double
counting of charged particles in the central detector and the 
calorimeters~\cite{MT}.

\subsection{Reconstruction of Four Jets}

We assigned the observed particles to jets, whose
directions reflect approximately the directions of the hard
partons presumed to be emitted before hadronization\footnote{ The Monte Carlo generator model  generates
first partons, then fragments them into particles and finally tracks
these particles through the detector. These three stages will be referred 
to in the text as parton level, particle level and
detector level respectively.}.
In this analysis we employ the DURHAM jet-finder \cite{durham}, 
which uses a jet resolution
 parameter between a pair of jets defined as
\begin{equation}
y_{ij} = \frac{\displaystyle 2}{\displaystyle E_{\mathrm{vis}}^2}\cdot 
\mbox{min}(E_i^2,E_j^2)(1-\cos\theta_{ij})
\end{equation}
where $E_i$,$E_j$ correspond to the energies of jets {\it{i}} and 
{\it{j}}, while $\theta_{ij}$
is the angle between them, and $E_{\mathrm{vis}}$ is the visible energy of the
whole event. 

The jet-finding algorithm combines iteratively each pair of particles 
{\it{k}} and {\it{l}} 
with 
the smallest $y_{kl}$ into a new pseudo-particle by adding the individual
four-momenta, until
exactly four pseudo-particles are reconstructed, which are defined to be 
four jets.
Then the minimum value of $y_{ij}$ of all
the possible combinations of the four jets is determined. This value is
called $y^{34}$ and is used to preclassify the events\footnote
{This definition of $y^{34}$ is not to be confused
with a jet resolution parameter $y_{34}$ limited to two individual jets
numbered three and four!}. 
If one stops the iteration of the jet-finding algorithm at five jets,
then the corresponding value of $y_{ij}$ is called $y^{45}$.
The usual way to choose a sample of four jet events is to specify
a fixed value of a parameter called $y_{\mathrm{cut}}$ and 
select events such that $y^{45}\;<\;y_{\mathrm{cut}}\;<\;y^{34}$.
The quantities $y^{45}$  and $y^{34}$ represent those
values of $y_{\mathrm{cut}}$ where the event makes a transition from a 
five-jet classification to a four-jet classification, and four-jet to three-jet, respectively.
This method of selecting four-jet events is
not sufficient to exclude five- and more parton events reconstructed
as four-jet events. 
We therefore accepted an  event if $y^{34}>0.012$, but demanded 
$y^{45}<0.006$. This eliminates a large fraction of unresolved five-parton
events in the sample, without reducing appreciably the number of four-jet
events (22\% for data but only 8\% for ERT, at the detector level )

From an analysis of 3.3$\times$10$^6$ accepted hadronic events
by OPAL  at the Z$^0$ peak we extracted 8.4$\times$10$^4$
four-jet events selected as described above.

A sample of 14.3$\times$10$^4$ simulated four-jet events was 
used to correct for
detector and acceptance effects. They  were generated with the ERT
matrix element calculation we want to test and subjected to the 
same cuts as the data.

\subsection{Measurement of the Bengtsson-Zerwas correlation}

We  used as angular variable mostly the Bengtsson-Zerwas \cite{BEZ} 
correlation, other correlations are also possible \cite{NAR}.
It measures the angle $\chi_{\mathrm{BZ}}$ between the two vectors
$\overrightarrow{p}_1\times\overrightarrow{p}_2$ and
$\overrightarrow{p}_3\times\overrightarrow{p}_4$. Here 
$\overrightarrow{p}_1$ to $\overrightarrow{p}_4$ are the 
momentum vectors of jets 1 to 4, where the numbering refers
to jets ordered with respect to their measured energy 
($E_1>E_2>E_3>E_4$). Ordering increases the probability that jets 1 and 2 
do indeed originate from the primary quarks as required.
To minimize the sensitivity to exchanges between jets 1 and 2 or 3 and 4,
we used the symmetrized version of the B-Z correlation

\begin{equation}
\chi_{\mathrm{BZ}} = \angle (\overrightarrow{p}_1\times\overrightarrow{p}_2)\;,\;(\overrightarrow{p}_3\times\overrightarrow{p}_4)\;;\;
\mathrm{if}\; \chi_{\mathrm{BZ}}>\frac{\pi}{2},\; \chi_{\mathrm{BZ}}=\pi - \chi_{\mathrm{BZ}} 
\end{equation}
so that effectively $\chi_{\mathrm{BZ}}$ runs from zero to $\frac{\pi}{2}$.
To obtain well defined planes,
the angles between jets 1 and 2, and between jets 3 and 4,
were required to be smaller than 160$^\circ$
The number of accepted events quoted above includes this cut.
The Bengtsson-Zerwas  correlation discriminates well 
between ${\mathrm{q\overline{q}q\overline{q}}}$ and 
${\mathrm{q\overline{q}gg}}$ events, 
since the polarization of the intermediate gluon \cite{OLSON} 
influences directly the distribution of its decay products,
either two quarks or two gluons.
It measures only angles and it relies
on jet energies only in so far as energy ordering is required.
The angular resolution, as obtained from comparing the ERT Monte Carlo
at the particle level with that at the detector level, was 14 degrees RMS.

\section{Results}

\begin{figure}[htb]
\parbox[thb]{7.0cm}{
\begin{center}
  \mbox{\epsfxsize=6.0cm\epsfysize=8.0cm\epsffile{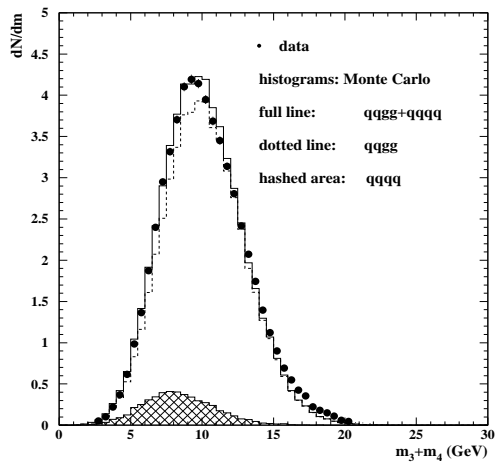}}
\end{center}}
\parbox[thb]{0.2cm}{$~~$}
\parbox[thb]{7.2cm}{
\caption{Comparison of the distributions of  $m_3+m_4$ 
between real data  and  ERT simulation (normalised with respect to each other).
An overall correction of 1 GeV has been applied to the scale of the ERT
 distribution in order to make the two distributions coincide better. 
The hashed histogram corresponds to the distribution of four-quark
events (qqqq), while the dotted curve represents the qqgg events.}}
\label{xx}
\end{figure}

The variable which we will use to subdivide the phace-space is 
 $m_3+m_4$, the sum of the intrinsic masses of the two least energetic jets.
Before proceeding to calculate the correlations and fit them to the
data, we have to assure ourselves that the raw distributions of the
events as a function of  $m_3+m_4$ are satisfactorily reproduced by 
the ERT simulation. Figure 1 shows that this is the case.

In figure 2a one can find the results for f$_q$ (full circles), 
obtained by adjusting f$_q$ (i.e. the relative proportions
of the theoretical ${\mathrm{q\overline{q}q\overline{q}}}$  and 
${\mathrm{q\overline{q}}}$gg correlations) to obtain a best 
fit to the experimental data\footnote{ Some bins in the region 
of high  $m_3+m_4$ in figure 2a of ref.\cite{CANJOUR} have been combined,
yielding the results displayed here }.
One sees that the data follow approximately the ERT predictions
(histogram) until 10 GeV, for values above 10 GeV there are
substantial differences.
\begin{figure}[htb]
\parbox[thb]{7.0cm}{
\begin{center}
  \mbox{\epsfxsize=7.0cm\epsfysize=9.0cm\epsffile{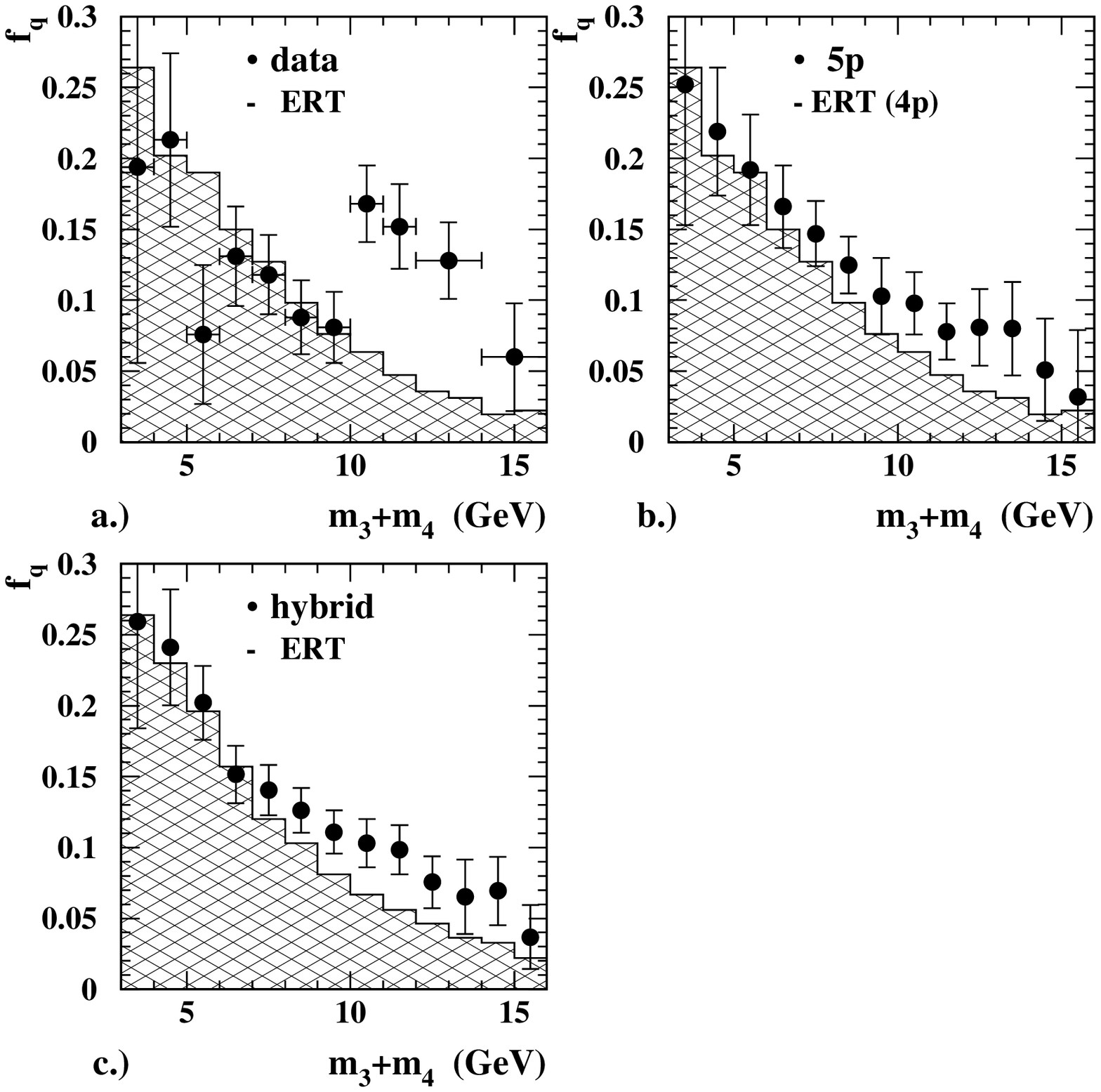}}
\end{center}}
\parbox[thb]{0.2cm}{$~~$}
\parbox[thb]{7.2cm}{
\caption{
a.) Comparison of the distribution of  $f_{\mathrm{q}}$ as a function of
$m_3+m_4$ for the data (full circles) with the  ERT prediction 
(histogram).\newline
b.) Comparison of the  distribution of $f_{\mathrm{q}}$  as a function of
$m_3+m_4$ for a mixture of 5 parton events (15\%) and 4 parton
events (85\%) (full circles)
with the ERT prediction (histogram). 
The proportions of the mixture have been adjusted so as to reproduce 
the global (integrated)
value of   $R_{4\mathrm{q}}$ = 1.32.\newline
c.) Comparison of the  distribution of $f_{\mathrm{q}}$  as a function of
$m_3+m_4$ for the hybrid model (full circles)
with the ERT prediction (histogram). 
Here also the proportion of the hybrid events has been adjusted so as to 
reproduce 
the global (integrated)
value of   $R_{4\mathrm{q}}$ = 1.32.
All errors are statistical only.}}
\label{hep_3fq}
\end{figure}

If we integrate over all events, the global fit yields an apparent fraction of 

f$_q^{exp}$ = 0.108$\pm0.009$ compared to a theoretical value of
f$_q^{th}$ = 0.082$\pm0.001$,

(errors statistical). 
 
A convenient way of comparing experiment to theory is to form the ratio
 
 $R_{4\mathrm{q}}$=$\frac{f_q^{exp}}{f_q^{theor}}$, which yields 

 $R_{4\mathrm{q}}$ = 1.32$\pm0.10$ ($\chi^2$/dgf=0.9) for all accepted events.
 The cited error being purely statistical, such a value of $R_{4\mathrm{q}}$ 
represents an approximate agreement between data and theory.

But if we divide the available phase-space into two regions, one with
$m_3+m_4< $10 GeV and the other with  $m_3+m_4>$ 10 GeV,
we find

for  $m_3+m_4<$ 10 GeV: $R_{4\mathrm{q}}$ = 0.88$\pm0.11$ 
($\chi^2$/dgf=0.24), while

for $m_3+m_4>$ 10 GeV: $R_{4\mathrm{q}}$ = 2.98$\pm0.35$ 
($\chi^2$/dgf=1.3), errors statistical only.

These rather large differences between the values of  
 $R_{4\mathrm{q}}$ for the two regions of phase-space constitute
the most important result of this analysis.

Let us now form the double quotient

Q =  $\frac{R_{4\mathrm{q}}((m_3+m_4)>10)}{R_{4\mathrm{q}}((m_3+m_4)<10)}$ = 3.39$\pm0.57$,

which is yet another convenient way to represent our results.

\begin{table}[ht]
\begin{center}
\begin{tabular}{|l|c|} \hline
\hline
cut  & $\Delta$ Q   \\ 
 &  \\ \hline\hline
a.) y$^{34}>0.015$ & $\pm$0.03  \\ \hline
b.) $|\cos\theta_{jet}|\leq 0.95$ & $-$0.19  \\ \hline
c.) number of tracks and clusters in each jet $>$ 5 & +0.52  \\ \hline
d.) $\chi_{\mathrm{BZ}} \geq 9$ degrees & $-$0.34  \\ \hline
e.) $\chi_{\mathrm{BZ}} \leq 81$ degrees & +0.18  \\ \hline
f.) y$^{45}<0.009$ & $\pm0.42$   \\ \hline
g.) $\theta_{34}\leq 140$ & $\pm0.15$    \\ \hline
h.) without correction of the ERT scale  & $-$0.05     \\ \hline
i.) configuration CH+EM, without hadronic clusters  &  +0.33  \\ \hline
k.) JADE algorithm \cite{JAD} instead of DURHAM  & -1.13   \\  \hline
l.) N-R correlation \cite{NAR} instead of B-Z & -0.07 \\ \hline
m.) $\theta_{34}$ distribution instead of B-Z & +0.10 \\ \hline
n.) constrained fit : $\sum E_i=E_{tot}$ ; $\sum \overrightarrow{p}_i 
= 0$ & -0.21 \\ \hline
o.)  statistical error & $\pm0.57$  \\ \hline\hline
total systematic error  & $^{+0.79}_{-1.30}$     \\
\hline
 total error  & $\pm1.2$     \\
\hline
\end{tabular}
\caption{  Systematical errors of the  quotient Q }
\label{err_sys}
\end{center}
\end{table}

An estimation of the sytematical errors of this quantity can be found
in table~\ref{err_sys}.
In this table we consider the variations of the following
quantities:\\
a.) increase of the jet resolution criterium, y$^{34}$, to 0.015;\\ 
b.) elimination of jets which point approximately in the direction of the
beam pipe;\\
c.) elimination 
of jets with few particles;\\ 
d.) and e.) elimination of small and large
angles for  $\chi_{\mathrm{BZ}}$;\\
f.) admission of a larger number of 5-jet
events;\\
g.) limitation on  $\theta_{34}$,
the angle between jet 3 et jet 4, to 140 degrees instead of 160;\\ 
h.) without the correction of 1 GeV for the Monte Carlo scale; \\
i.) jets calculated without using hadronic clusters and without correction 
for double counting of energy; \\
k) using the old 
JADE algorithm (see ref.~\cite{JAD}) with less jet finding resolution;\\ 
l.) replace  $\chi_{\mathrm{BZ}}$ by the
Nachtmann-Reiter correlation (see ref.~\cite{NAR});\\ 
m) replace   $\chi_{\mathrm{BZ}}$
by $\theta_{34}$;\\
n.) fit with constraint by imposing conservation
of total energy and momentum. 

One finds:

Q = 3.39$_{-1.42}^{+0.97}$, so that  Q becomes approximately 

Q = 3.4$\pm1.2$.

The difference between experiment and theory, expressed by the value
of Q$>$1, is approximately an effect of two standard deviations. 

\begin{figure}[htb]
\parbox[thb]{7.0cm}{
\begin{center}
  \mbox{\epsfxsize=7.0cm\epsfysize=9.0cm\epsffile{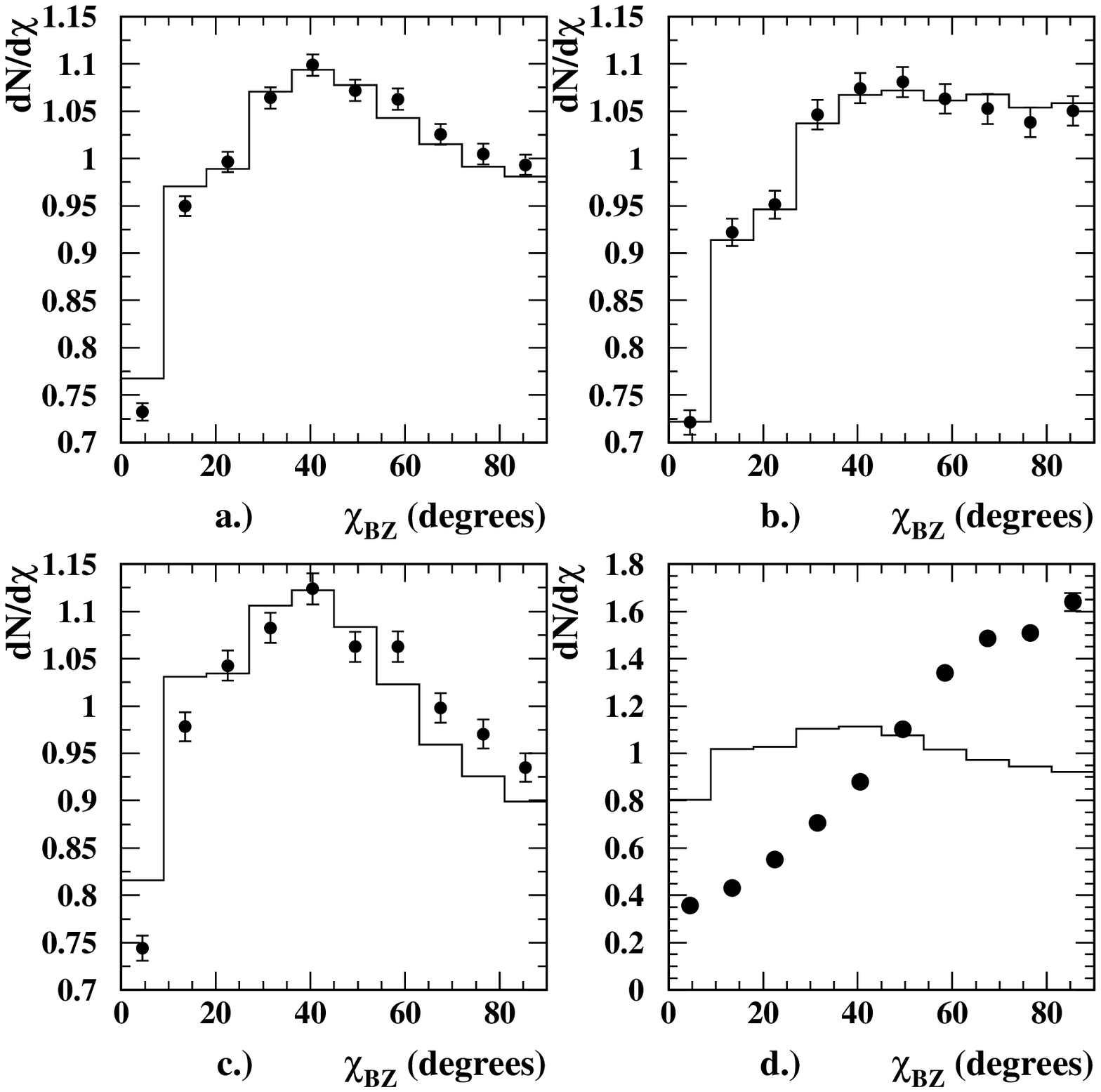}}
\end{center}}
\parbox[thb]{0.2cm}{$~~$}
\parbox[thb]{7.2cm}{
\caption{
a.) Comparison of the distribution of $\chi_{\mathrm{BZ}}$ 
between data (full circles) and ERT prediction (histogram)
for all accepted events (before fitting for the best value of $f_q$).\newline
b.) Comparison of the distribution of $\chi_{\mathrm{BZ}}$ 
between data (full circles) and ERT prediction (histogram)
for events with  $m_3+m_4<10$ GeV.\newline
c.) Comparison of the distribution of $\chi_{\mathrm{BZ}}$ 
between data (full circles) and ERT prediction (histogram)
for events with  $m_3+m_4>10$ GeV.\newline
d.) Comparison  of the  $\chi_{\mathrm{BZ}}$ distribution for the
four-quark events (full circles) with the one for two quark and two
gluon events (histogram), as calculated with the ERT prediction
for all accepted events. 
Note the difference in scale for this last graph compared
to the three previous ones.\newline
All the distributions are normalised with repect to each other,
they are not corrected for detector effects and all errors
are only statistical.
}}
\label{hep_chibz}
\end{figure}

In figure~\ref{hep_chibz} we display directly the 
$\chi_{\mathrm{BZ}}$ distributions which yielded
the above-mentioned results.
In figure~\ref{hep_chibz}a the experimental and theoretical (leading order)
distributions are shown for all accepted events,
while in figures~\ref{hep_chibz}b and c these distributions
are displayed  for events with  $m_3+m_4>10$ GeV and $m_3+m_4>10$ GeV
respectively.
In these three cases the difference between experiment and theory
can be expressed by the values  
for the chi-squares per degree of freedom, which are  1.8, 0.24, and 4.9 
respectively.
 Visual inspection
confirms that the difference between experiment and theory is largest
for case ``c'', $m_3+m_4>10$ GeV. After adjustment by fit ( the
corresponding distributions are not shown here ),
the chi-squares become 0.9, 0.1 et 1.3 respectively.
Figure~\ref{hep_chibz}d is an illustration of the theoretical shape differences
between ${\mathrm{q\overline{q}q\overline{q}}}$ and 
${\mathrm{q\overline{q}}}$gg events (for all accepted events according to ERT),
which make our type of analysis possible.

\section{ Discussion }

Figure 2a shows that there are deviations of the experimental distributions
from the theoretical ones which concentrate in the region of high values of
the variable $m_3+m_4$. 
Such deviations could also be caused by the absence of higher order effects in 
the ERT simulation. To get an idea of the influence such higher order effects
might have, we show in figure 2b the results one obtains if one replaces the experimental distributions by distributions explicitly containing five-parton events in addition to the 4-parton events furnished by ERT. 
The mixture is adjusted so as to reproduce the observed value of
1.32 for $R_{4\mathrm{q}}$.
In figure 2c one finds the results for a hybrid calculation, where
each of the four partons from ERT is followed by a parton shower, 
before being fragmented into particles. 
These events are also mixed with regular four-parton events so as to 
reproduce  the observed value of $R_{4\mathrm{q}}$.
In both cases one expects such calculations to show trends associated
with higher order effects.
In figures 2b and 2c these calculations are represented as fictitious data 
(full circles) to be compared with calculations where such effects
are not present (histogram).
One notices that indeed such higher order effects have a tendency
to increase with the value of $m_3+m_4$,
but more gradually than the actually observed experimental results
of figure 2a.
Another possibility to explain such results could be the appearance
of new fermions in the region of  10 to 15 GeV for  $m_3+m_4$, i.e. 
5  to  7 GeV  for each of the emitted particles.
(see f.ex. refs.~\cite{BERGER1,BERGER2}). 
At this stage of the analysis it is not possible to discriminate
between these possibilities, but it is hoped that this research note
motivates other groups to verify these results with more precision.

\section{Summary}
The Bengtsson-Zerwas~\cite{BEZ} angular correlation, calculated with
the second order  ERT~\cite{ERT} simulation, has been compared with
the correlation measured for hadronic four-jet events from the disintegration
of the Z$^0$. Shape differences between experiment and theory have been
observed. The differences concentrate in the region of   $m_3+m_4>$ 10 GeV,
where  $m_3+m_4$ is the sum of the intrinsic masses of the two least energetic
jets
\bibliographystyle{unsrt}

\begin{thebibliography}{99}
\bibitem{CANJOUR}
H.Jeremie, P.Leblanc and E.Lefebvre,\\   
Can. Jour. Phys. {\bf{84}} (2006) 411.
\bibitem{GLUIN}
G.Farrar, Phys.Lett. {\bf{B 265}} (1991) 395.
\bibitem{ERT}
R.K.Ellis, D.A. Ross and A.E. Terrano, Nucl. Phys. {\bf {B 178}} (1981) 421.
\bibitem{JETST}
T.Sj\"ostrand, Comp. Phys. Comm. {\bf{39}} (1986) 347;\newline
T.Sj\"ostrand and M.Bengtsson, Computer Phys. Comm. {\bf{43}} (1987) 367.
\bibitem{EECORR}
OPAL collaboration, P.D.Acton et al., Phys. Lett.{\bf{B 276}} (1992) 547.
\bibitem{ANDRE}
J.Andr\'e and T.Sj\"ostrand, Phys. Rev. {\bf{D 57}} (1998) 5767.
\bibitem{ZEPP}
K.Hagiwara and D.Zeppenfeld, Nucl. Phys. {\bf{B 313}} (1989) 560;\\
F.W\"ackerle, Diplomarbeit Karlsruhe 1994, IEKP-KA/93-19.
\bibitem{ALEPH}
ALEPH collaboration, D.Decamp et al., Phys. Lett. {\bf{B 284}} (1992) 151.
\bibitem{DELPHI}
DELPHI collaboration, P.Abreu et al.,Phys. Lett. {\bf{B 255}} (1991) 466;\\
DELPHI collaboration, P.Abreu et al., Zeit. f. Phys. {\bf{C 59}} (1993) 357.
\bibitem{DELPHI2}
DELPHI collaboration, P.Abreu et al., Phys. Lett. {\bf{B 414}} (1997) 401.
\bibitem{opevtr}
 OPAL collaboration, R. Akers et al., Zeit. f. Phys. {\bf{C 68}} (1995) 519.
\bibitem{OPAL1}
OPAL collaboration, R. Akers et al., Zeit. f. Phys. {\bf{C 65}} (1995) 367.
\bibitem{ALEPH2} 
ALEPH collaboration, R.Barate et al., Zeit. f. Phys. {\bf{C 76}} (1997) 1.
\bibitem{OPAL2}
OPAL collaboration, G.Abbiendi et al., Eur. Phys. J. {\bf{C 20}} (2001) 601.
\bibitem{ALEPH3} 
ALEPH collaboration, R.Barate et al.,  Eur. Phys. J. {\bf{C 27}} (2003) 1.
\bibitem{OPALDET}
OPAL collaboration, K.Ahmet et al., Nucl. Instr. and Meth. {\bf{A 305}} (1991) 275.
\bibitem{XX}
M. Arignon  et al., Nucl. Instr. and Meth. {\bf{A 313}} (1992) 103.
\bibitem{YY}
D. Charlton et al., Nucl. Instr. and Meth. {\bf{A 325}} (1993) 129.
\bibitem{ZZ}
OPAL collaboration, G.Alexander et al., Zeit. f. Phys. {\bf{C 52}} (1991) 175.
\bibitem{MT}
OPAL collaboration, K.Ackerstaff et al., Eur. Phys. J. {\bf{C 2}} (1998) 213
\bibitem{durham}
S. Catani, Y. L. Dokshitzer, M.Olsson, G.Turnock and B.R.Webber,
Phys. Lett. {\bf{B 269}} (1991) 432.
\bibitem{BEZ}
M.Bengtsson and P.Zerwas, Phys. Lett. {\bf{B 208}} (1988) 306.
\bibitem{NAR}
O.Nachtmann and A.Reiter, Zeit. f. Phys. {\bf{C 16}} (1982) 45.
\bibitem{OLSON}
H.A. Olson, P. Osland, I. Overbo, Phys. Lett. {\bf{B 89}} (1980) 221.
\bibitem{JAD}                                         
JADE collaboration, S.Bethke et al., Phys. Lett. {\bf{B 213}} (1988) 235.
\bibitem{BERGER1}
E.L.Berger al., Phys. Rev. Lett. {\bf{86}} (2001) 4231
\bibitem{BERGER2}
E.L.Berger et al., Phys. Rev. {\bf{D71}} (2005) 14007
\end{thebibliography}

\clearpage

\end{document}